\newcommand{\nrBugsmysql}{25}
\newcommand{\nrBugspostgres}{9}
\newcommand{\nrBugssqlite}{65}
\newcommand{\mysqlclosedduplicate}{4}
\newcommand{\mysqlclosednotabug}{1}

\newcommand{\mysqlfixedInDocsOrCode}{15}
\newcommand{\mysqlverified}{10}
\newcommand{\postgresclosedduplicate}{6}
\newcommand{\postgresclosednotabug}{7}

\newcommand{\postgresfixedInDocsOrCode}{5}
\newcommand{\postgresverified}{4}
\newcommand{\sqliteclosedduplicate}{2}
\newcommand{\sqliteclosednotabug}{4}

\newcommand{\sqlitefixedInDocsOrCode}{65}
\newcommand{\sqliteverified}{0}
\newcommand{\closedduplicate}{12}
\newcommand{\closednotabug}{12}
\newcommand{\fixed}{77}
\newcommand{\fixedindocumentation}{8}
\newcommand{\verified}{14}
\newcommand{\mysqlcontains}{14}
\newcommand{\mysqlerror}{10}
\newcommand{\mysqlsegfault}{1}
\newcommand{\postgrescontains}{1}
\newcommand{\postgreserror}{7}
\newcommand{\postgressegfault}{1}
\newcommand{\sqlitecontains}{46}
\newcommand{\sqliteerror}{17}
\newcommand{\sqlitesegfault}{2}
\newcommand{\sumcontains}{61}
\newcommand{\sumerror}{34}
\newcommand{\sumsegfault}{4}
\newcommand{\nrTruePositives}{99}
\newcommand{\nrFalsePositives}{24}
\newcommand{\nrReported}{123}

\newcommand{\sqliteCOLLATE}{11}
\newcommand{\sqliteCORRUPTEDDB}{4}
\newcommand{\sqliteINDEX}{17}

\newcommand{\sqliteLANGUAGEDEFICIENCY}{2}

\newcommand{\sqliteLIKEOPTIMIZATION}{4}

\newcommand{\sqliteUNEXPECTEDTYPE}{8}
\newcommand{\sqliteWITHOUTROWID}{5}

\newcommand{\postgresMULTITHREADEDwithFP}{4}

\newcommand{\sqliteREINDEXCONSTRAINTFAILEDwithFP}{6}

\newcommand{\maxloc}{8}
\newcommand{\avgloc}{3.71}
\newcommand{\nrtestcaseswithsingleline}{13}
\newcommand{\totalnrbugs}{99}

\newcommand{\sqliteNrCriticalBugs}{14}
\newcommand{\sqliteNrImportantBugs}{14}

\newcommand{\sqliteNrNoneBugs}{13}
\newcommand{\sqliteNrSevereBugs}{8}

\newcommand{\percOneCreateTable}{90.0}

\newcommand{\mySQLNrEngineBugs}{5}
\newcommand{\mySQLNrUnsignedBugs}{4}
\newcommand{\percUNIQUE}{22.2}
\newcommand{\percPRIMARYKEY}{17.2}
\newcommand{\percINDEX}{28.3}
\newcommand{\percFOREIGNKEY}{1.0}

\newcommand{\locPostgres}{4981}
\newcommand{\locMysql}{3995}
\newcommand{\locSqlite}{6501}

\newcommand{\locCommon}{918}
\newcommand{\locDBMSPostgres}{329999}
\newcommand{\locDBMSSqlite}{49703}
\newcommand{\locDBMSMySQL}{707803}
\newcommand{\implPercPostgres}{1.5}
\newcommand{\implPercSqlite}{13.1}
\newcommand{\implPercMysql}{0.6}

\documentclass{vldb}

\usepackage{graphicx}

\usepackage{colortbl}
\usepackage{capt-of}

\usepackage[]{algorithm2e}
\usepackage{booktabs}
\usepackage{xcolor,listings}
\usepackage{array}
\usepackage{amssymb}
\usepackage{pifont}
\usepackage{numprint}
\npthousandsep{,}

\usepackage[originalparameters]{ragged2e}  
\usepackage{tikz}
\newcommand*\circled[1]{\tikz[baseline=(char.base)]{
            \node[shape=circle,draw,inner sep=1pt] (char) {#1};}}

\usepackage[]{collab} 
\collabAuthor{mr}{blue}{Manuel Rigger}
\collabAuthor{zs}{purple}{Zhendong Su}

\usepackage{balance}

\vldbTitle{Testing Database Engines via \approach{}}
\vldbAuthors{Manuel Rigger and Zhendong Su}
\vldbDOI{https://doi.org/10.14778/xxxxxxx.xxxxxxx}
\vldbVolume{12}
\vldbNumber{xxx}
\vldbYear{2019}

\usepackage{textcomp}

\newcommand{\approach}{Pivoted Query Synthesis}
\newcommand{\containsoracle}{containment oracle}
\newcommand{\errororacle}{error oracle}
\newcommand{\toolname}{SQLancer}
\newcommand{\sqlite}{SQLite}
\newcommand{\postgres}{PostgreSQL}
\newcommand{\sqlsmith}{SQLsmith}
\newcommand{\ie}{\emph{i.e.}}
\newcommand{\eg}{\emph{e.g.}}
\newcommand{\pivot}{pivot row}
\usepackage{url}
\newcommand{\bugurl}{\url{https://www.manuelrigger.at/pqs}}

\newcommand{\normalpar}[1]{\textbf{#1.}}
\newcommand{\parspaceskip}{\vspace{1em}}

\newcommand{\tablesize}{\small{}}
\newcommand{\reducespace}{}

\makeatletter
\def\@copyrightspace{\relax}
\makeatother

\makeatletter
\def\@mkbibcitation{\relax}
\makeatother

\begin{document}


\title{Testing Database Engines via \approach{}}



%
%
%
%

\numberofauthors{2} 

\author{
%
%
\alignauthor
Manuel Rigger\\
       \affaddr{Department of Computer Science}\\
       \affaddr{ETH Zurich}\\
       \affaddr{Zurich, Switzerland}\\
       \email{manuel.rigger@inf.ethz.ch}
\alignauthor
Zhendong Su\\
       \affaddr{Department of Computer Science}\\
       \affaddr{ETH Zurich}\\
       \affaddr{Zurich, Switzerland}\\
       \email{zhendong.su@inf.ethz.ch}
}

\maketitle

\begin{abstract}
\sloppy{}
	Relational databases are used ubiquitously.
	They are managed by database management systems (DBMS), which allow inserting, modifying, and querying data using a domain-specific language called Structured Query Language (SQL).
	Popular DBMS have been extensively tested by fuzzers, which have been successful in finding crash bugs.
	However, approaches to finding \emph{logic bugs}, such as when a DBMS computes an incorrect result set, have remained mostly untackled.
	Differential testing is an effective technique to test systems that support a common language by comparing the outputs of these systems.
	However, this technique is ineffective for DBMS, because each DBMS typically supports its own SQL dialect.
	To this end, we devised a novel and general approach that we have termed \emph{\approach{}}.
	The core idea of this approach is to automatically generate queries for which we ensure that they fetch a specific, randomly selected row, called the \emph{\pivot{}}.
	If the DBMS fails to fetch the \pivot{}, the likely cause is a bug in the DBMS.
	We tested our approach on three widely-used and mature DBMS, namely \sqlite{}, MySQL, and \postgres{}.
	In total, we reported \nrReported{} bugs in these DBMS, \nrTruePositives{} of which have been fixed or verified, demonstrating that the approach is highly effective and general.
	We expect that the wide applicability and simplicity of our approach will enable the improvement of robustness of many DBMS.
\end{abstract}

\section{Introduction}

Database management systems (DBMS) based on the relational model~\cite{Codd1970} are a central component in many applications, since they allow efficiently storing and retrieving data.
They have been extensively tested by random query generators such as \sqlsmith{}~\cite{sqlsmith}, which have been effective in finding queries that cause the DBMS process to crash (\ie{}, by causing a buffer overflow).
Also fuzzers such as AFL~\cite{afl} are routinely applied to DBMS.
However, these approaches cannot detect \emph{logic bugs}, which we define as bugs that cause a query to return an incorrect result, for example, by erroneously omitting a row, without crashing the DBMS.


Logic bugs in DBMS are difficult to detect automatically.
A key challenge for automatic testing is to come up with an effective \emph{test oracle}, that can detect whether a system behaves correctly for a given input~\cite{Howden:1978}.
In 1998, Slutz proposed to use \emph{differential testing}~\cite{mckeeman1998differential} to detect logic bugs in DBMS, by constructing a test oracle that compares the results of a query on multiple DBMS, which he implemented in a tool RAGS~\cite{slutz1998massive}.
While RAGS detected many bugs, differential testing comes with the significant limitation that the systems under test need to implement the same semantics for a given input.
All DBMS support a common and standardized language \emph{Structured Query Language (SQL)} to create, access, and modify data~\cite{Chamberlin:1974}.
In practice, however, each DBMS provides a plethora of extensions to this standard and deviates from it in other parts (\eg{}, in how \texttt{NULL} values are handled~\cite{slutz1998massive}).
This vastly limits differential testing, and also the author stated that the small common core and the differences between different DBMS were a challenge~\cite{slutz1998massive}.
Furthermore, even when all DBMS fetch the same rows, it cannot be ensured that they work correctly, because they might be affected by the same underlying bug.

In order to efficiently detect logic bugs in DBMS, we propose a general and principled approach that we termed \emph{\approach{}}, which we implemented in a tool called \toolname{}.
The core idea is to solve the oracle problem for a single, randomly-selected row, called the \emph{\pivot{}}, by synthesizing a query whose result set must contain the \pivot{}.
By considering only a single row, our approach is simple to understand and implement.
We synthesize the query by randomly generating expressions for \texttt{WHERE} and \texttt{JOIN} clauses, evaluating the expressions based on the \pivot{}, and modifying each expression to yield \texttt{TRUE}.
If the query, when processed by the DBMS, fails to fetch the \pivot{}, a bug in the DBMS has been detected.
We refer to this oracle as the \emph{containment} oracle.

Listing~\ref{lst:illustrativex} illustrates our approach on a test case that triggers a bug that we found using the \containsoracle{} in the widely-used DBMS \sqlite{}.
The \texttt{CREATE TABLE} statement creates a new table \texttt{t0} with a column \texttt{c0}.
Subsequently, an index is created and four rows with the values \texttt{0}, \texttt{1}, \texttt{2}, \texttt{3}, and \texttt{NULL} are inserted.
We select the \pivot{} \texttt{c0=NULL} and construct the random \texttt{WHERE} clause \texttt{c0 IS NOT 1}.
Since \texttt{NULL IS NOT 1} evaluates to \texttt{TRUE}, we can directly pass the query to the DBMS, expecting the row with value \texttt{NULL} to be contained in the result.
However, due to a logic bug in the DBMS, the partial index was used based on the incorrect assumption that \texttt{c0 IS NOT 1} implied \texttt{c0 NOT NULL}, resulting in the \pivot{} not being fetched.
We reported this bug to the \sqlite{} developers, who stated that it existed since 2013, classified it as critical and fixed it shortly after we reported it.
Even for this simple query, differential testing would have been ineffective in detecting the bug.
The \texttt{CREATE TABLE} statement is specific to \sqlite{}, since, unlike other popular DBMS, such as \postgres{} and MySQL, \sqlite{} does not require the column \texttt{c0} to be assigned a column type.
Furthermore, both MySQL and \postgres{} lack an operator \texttt{IS NOT} that can be applied to integers.
All DBMS provide an operator \texttt{IS NOT TRUE}, which, however, has different semantics; for \sqlite{}, it would fetch only the value \texttt{0}, and not expose the bug.

\reducespace{}
\begin{lstlisting}[caption={Illustrative example, based on a \emph{critical} \sqlite{} bug that we reported.}, label=lst:illustrativex]
CREATE TABLE t0(c0);
CREATE INDEX i0 ON t0(1) WHERE c0 NOT NULL;
INSERT INTO t0(c0) VALUES (0), (1), (2), (3),
	(NULL);
SELECT c0 FROM t0 WHERE t0.c0 IS NOT 1; -- unexpected: NULL is not contained
\end{lstlisting}



\sloppy{}
To demonstrate the generality of our approach, we implemented it for three popular and widely-used DBMS, namely \sqlite{}~\cite{sqlite}, MySQL~\cite{mysql}, and \postgres{}~\cite{postgres}.
In total, we found \nrTruePositives{} bugs, namely \nrBugssqlite{} bugs in \sqlite{}, \nrBugsmysql{} bugs in MySQL, and \nrBugspostgres{} in \postgres{}, demonstrating that the approach is highly effective and general.
Out of these bugs, we found \sumcontains{} bugs with the \containsoracle{}.
We found \sumerror{} bugs by causing DBMS-internal errors, such as database corruptions, and for \sumsegfault{} bugs we caused DBMS crashes (\ie{}, \texttt{SEGFAULT}s).
One of the crashes that we reported for MySQL was classified as a security vulnerability (CVE-2019-2879). 
Detailed information on the bug reports and fixes can be found at \bugurl{}.
We designed our approach to mainly detect logic bugs that cannot be found by fuzzers, which is confirmed by the evaluation.
Since our method is general and applicable to all DBMS, we expect that it will be widely adopted to detect bugs that have so far been overlooked.
In summary, we contribute the following:
\begin{itemize}
	\item A general and highly-effective approach to finding bugs in DBMS termed \emph{\approach{}}.
	\item An implementation of this approach in a tool named \toolname{}, used to test \sqlite{}, MySQL, and \postgres{}.
	\item An evaluation of \toolname{}, which uncovered a total of \totalnrbugs{} bugs.
\end{itemize}

\section{Background}

This section provides important background information on relational DBMS, SQL, the DBMS we tested, and their characteristics.

\parspaceskip{}

\sloppy{}
\normalpar{Database management systems}
DBMS are based on a \emph{data model}, which abstractly describes how data is organized.
In our work, we primarily aim to test DBMS based on the \emph{relational data model} proposed by Codd~\cite{Codd1970}, on which most widely-used databases, such as Oracle, Microsoft SQL, \postgres{}, MySQL, and \sqlite{} are based.
A relation $R$ in this model is a mathematical relation $R \subseteq S_1 \times S_2 \times ... \times S_n$ where $S_1$, $S_2$, ..., $S_n$ are referred to as domains.
More commonly, a relation is referred to as a \emph{table} and a domain is referred to as a \emph{data type}.
Each tuple in this relation is referred to as a row.
Note that rows in $R$ are unordered.
While the original relational model did not allow duplicate rows, most DBMS use bags of tuples, which allow duplicate values.
The approach we present in Section~\ref{sec:approach} works for both sets of tuples and bags of tuples.
\parspaceskip{}



\normalpar{Structured query language (SQL)}
Structured Query Language (SQL)~\cite{Chamberlin:1974}, which is based on relational algebra~\cite{codd1972relational}, is the most commonly used language in DBMS to create tables, insert rows, and manipulate and retrieve data.
ANSI first standardized SQL in 1987, and it has since been developed further.
In practice, however, DBMS lack functionality described by the SQL standard and deviate from it, making it difficult to test DBMS using differential testing.
%
%
SQL statements can be roughly categorized as belonging to the Data Definition Language (DDL), Data Manipulation Language (DML), and Data Query Language (DQL).
DDL statements allow creating, changing, or removing elements such as tables or indexes in a database.
For example, \texttt{CREATE TABLE} allows creating a new table in database with a given \emph{schema} that defines the columns, their data types, and constraints.
\texttt{CREATE INDEX} creates an index, which is a supplementary data structure used to improve the speed of querying data.
\texttt{ALTER TABLE} can be used to change a table's schema.
\texttt{DROP} statements allow removing elements such as tables or indexes.
DML statements allow adding, changing, or removing data.
For example, \texttt{INSERT} inserts data into tables, \texttt{UPDATE} allows changing values in existing rows, and \texttt{DELETE} removes rows from a table.
The Data Query Language (DQL) allows fetching rows from a database using the \texttt{SELECT} statement.
Although these statements are supported by all the DBMS that we investigated, the syntax and semantics of these statements depend significantly on the respective DBMS.
In fact, many bugs that we found were caused by the implementation of features unique to the respective DBMS.
\parspaceskip{}

\normalpar{Important DBMS}
We focused on three popular and widely-used open-source DBMS: \sqlite{}, MySQL, and \postgres{} (see Table~\ref{tbl:dbmsoverview}).
According to the DB-Engines Ranking~\cite{dbengines} and the Stack Overflow's annual Developer Survey~\cite{stackoverflowsurvey}, these DBMS are among the most popular and widely-used DBMS.
Furthermore, the \sqlite{} website speculates that \sqlite{} is likely used more than all other databases combined; most mobile phones extensively use \sqlite{}, it is used in most popular web browsers, and many embedded systems (such as television sets).\footnote{https://www.sqlite.org/mostdeployed.html}
All DBMS have been maintained and developed for about 20 years.
\sqlite{} is developed by only three developers, MySQL is mainly developed commercially (by Oracle), and \postgres{} is developed by volunteers coordinated by five people who form the core team.
\parspaceskip{}



\begin{table}
\tablesize{}
\caption{The DBMS we tested are popular, complex, and have been developed for a long time.}
\label{tbl:dbmsoverview}
\begin{tabular}{l >{\RaggedLeft}p{0,8cm} >{\RaggedLeft}p{1,1cm} >{\RaggedLeft}p{0,5cm} >{\RaggedLeft}p{1cm} >{\RaggedLeft}p{1,0cm} }
\toprule{}
& \multicolumn{2}{c}{Popularity Rank} & & \\
\cmidrule(r){2-3}
DBMS & DB-Engines & Stack Overflow & LOC & Released & Age (years) \\
\midrule
\sqlite{} & 11 & 4 & 0.3M & 2000 & 19 \\ 
MySQL & 2 & 1 & 3.8M & 1995 & 24 \\ 
\postgres{} & 4 & 2 & 1.4M & 1996 & 23 \\ 
\bottomrule
\end{tabular}
\reducespace{}

\end{table}

\normalpar{Unique Features of the DBMS}
Each DBMS provides its own distinct set of features and characteristics.
\sqlite{} runs in the application process, and thus is mostly used for local data storage for individual applications and devices. 
It provides a limited set of language constructs since it strives to be compact.
However, due to \sqlite{}'s type-related features, we perceived it to be the most flexible DBMS.
For example, unlike other DBMS, \sqlite{} allows data types to be omitted for columns and performs implicit conversions when a value does not have the expected data type.
MySQL and \postgres{} are conventional DBMS where the server runs in its own process and is accessed by a client over a network connection.
Both DBMS provide significantly more features than \sqlite{}; for example, they provide high-level data types such as arrays and json files.
Examples of unique features to MySQL include unsigned data types, and that one of various non-standard storage engines can be assigned to a table (\eg{}, a CSV-based engine).
Distinct \postgres{} features  include object-oriented tables using \emph{table inheritance}, and that it performs only few implicit conversion.
\parspaceskip{}

\normalpar{Test Oracles}
Having an effective \emph{test oracle} is crucial for automatic testing approaches~\cite{Howden:1978}.
A test oracle can determine for a given test case whether the test case has passed or failed.
Manually written test cases encode the programmer's knowledge who thus acts a test oracle.
In this work, we are only interested in automatic test oracles, which would allow comprehensively testing a DBMS.
The most successful automatic test oracle for DBMS is based on \emph{differential testing}~\cite{slutz1998massive}.
Differential testing refers to a technique where a single input is passed to multiple systems that implement the same language to detect mismatching outputs, which would indicate a bug.
In the context of DBMS, the input corresponds to a database as well as a query, and the systems to multiple DBMS---when their fetched result sets mismatch, a bug in the DBMS would be detected.
However, as argued above, DBMS provide different features, making it difficult to use differential testing effectively.
Furthermore, differential testing is not an \emph{precise} oracle, as it fails to detect bugs that are shared by all the systems.
\parspaceskip{}

\section{\approach{}}
\label{sec:approach}
We propose \emph{\approach{}} as an automatic testing technique for detecting logic bugs in DBMS.
Our core insight is that verifying the correctness of the DBMS one row at a time is simpler than checking the complete result set, and enables creating a simple test oracle.
Specifically, our idea is to select a random row, to which we refer as the \pivot{}, from a set of tables (and views) in the database.
For this pivot row, we semi-randomly generate a set of expressions for which we ensure that they evaluate to \texttt{TRUE} based on an Abstract Syntax Tree (AST) interpreter.
By using these expressions in \texttt{WHERE} and \texttt{JOIN} clauses of an otherwise randomly-generated query, we can ensure that the \pivot{} must be contained in the result set.
If it is not contained, a bug has been found.
By repeatedly checking a single row, we speculate that this technique is similarly effective as one that verifies the correctness of the complete result set.
Basing the approach on an AST interpreter provides us with an exact oracle.
While implementing this interpreter requires moderate implementation effort for complex operators (such as regular expression operators), other challenges that a DBMS has to tackle, such as query planning, concurrent access, integrity, and persistence can be disregarded by it.
Furthermore, the AST interpreter can be naively implemented without affecting the tool's performance, since it only operates on a single record, whereas the DBMS has to potentially scan through all the rows of a database to process a query.

\subsection{Approach Overview}

Figure~\ref{fig:overview} illustrates the detailed steps of our approach.
First, we create a database with one or multiple random tables, which we fill with random data (see step \circled{1}).
We ensure that each table holds at least one row.
We then select a random row from each of the tables (see step \circled{2}), to which we refer as the \pivot{}.
We verify the correctness of the DBMS based on this \pivot{}, and also provide a test oracle. 
We randomly create expressions based on the DBMS' SQL grammar and valid table column names (see step \circled{3}).
We evaluate these expressions, substituting column references by the corresponding values of the \pivot{}.
Then, we modify the expressions so that they yield \texttt{TRUE} (see step \circled{4}).
We use these expressions in \texttt{WHERE} and/or \texttt{JOIN} clauses for a query that we construct (see step \circled{5}).
We pass this query to the DBMS, which returns a result set (see step \circled{6}), which we expect to contain the \pivot{}, potentially among other rows.
In a final step, we check whether the \pivot{} is indeed contained in the result set (see step \circled{7}).
If it is not contained, we have likely detected a bug in the DBMS.
For the next iteration, we either continue with step \circled{2} and generate new queries for a newly-selected \pivot{}, or continue with \circled{1} to generate a new database.

Our core idea is given by how we construct the test oracle, which is given by steps \circled{2} to \circled{7}.
Thus, Section~\ref{sec:qgeneration} first explains how we generate queries and check for containment, assuming that the database has already been created.
Section~\ref{sec:dbge} then explains step \circled{1}, namely how we generate the tables and data.
While working on the database generation, we found another applicable test oracle to detect bugs, namely by checking for unexpected errors returned by the DBMS.
We refer to this oracle as \emph{\errororacle{}} and also explain it in Section~\ref{sec:dbge}.
Section~\ref{sec:impl} provides important implementation details.


\subsection{Query Generation \& Checking}
\label{sec:qgeneration}
The core idea of our approach is to construct a query for which we anticipate that the \pivot{} is contained in the result set.
We randomly generate expressions to be used in a condition of the query, and ensure that each expression evaluates to \texttt{TRUE} for the \pivot{}.
This subsection describes how we generate random expressions that we rectify and then use in a query (\ie{}, steps \circled{3} to \circled{5}).
\parspaceskip{}

\normalpar{Random Condition Generation}
In step \circled{3}, we randomly generate Abstract Syntax Trees (ASTs) up to a specified maximum depth by constructing a random expression tree based on the database's schema (\ie{}, the column names and types).
Generating these expression trees is implemented similarly to RAGS~\cite{slutz1998massive} and \sqlsmith{}~\cite{sqlsmith}.
However, while these systems directly use the generated expressions in query conditions (\eg{}, in a \texttt{WHERE} clause), our approach requires the conditions to yield \texttt{TRUE} for the \pivot{}, which is ensured in the subsequent steps.
For \sqlite{} and MySQL, \toolname{} generates expressions of any type, because they provide implicit conversions to boolean.
For \postgres{}, which performs few implicit conversions, the generated root node must produce a boolean value, which we achieve by selecting one of the appropriate operators (\eg{}, a comparison operator).
\parspaceskip{}

Listing~\ref{alg:astgen} illustrates how generating the expressions is implemented for MySQL and \sqlite{}.
The input parameter \texttt{depth} ensures that when a specified maximum depth is reached, a leaf node is generated.
The leaf node can either be a randomly-generated constant, or a reference to a column in a table or view.
If the maximum depth is not yet reached, also other operators are considered (e.g., a unary operator such as \texttt{NOT}).
Note that generating these expressions is dependent on which operators the respective DBMS supports.

\begin{figure*}
	\includegraphics[width=\textwidth]{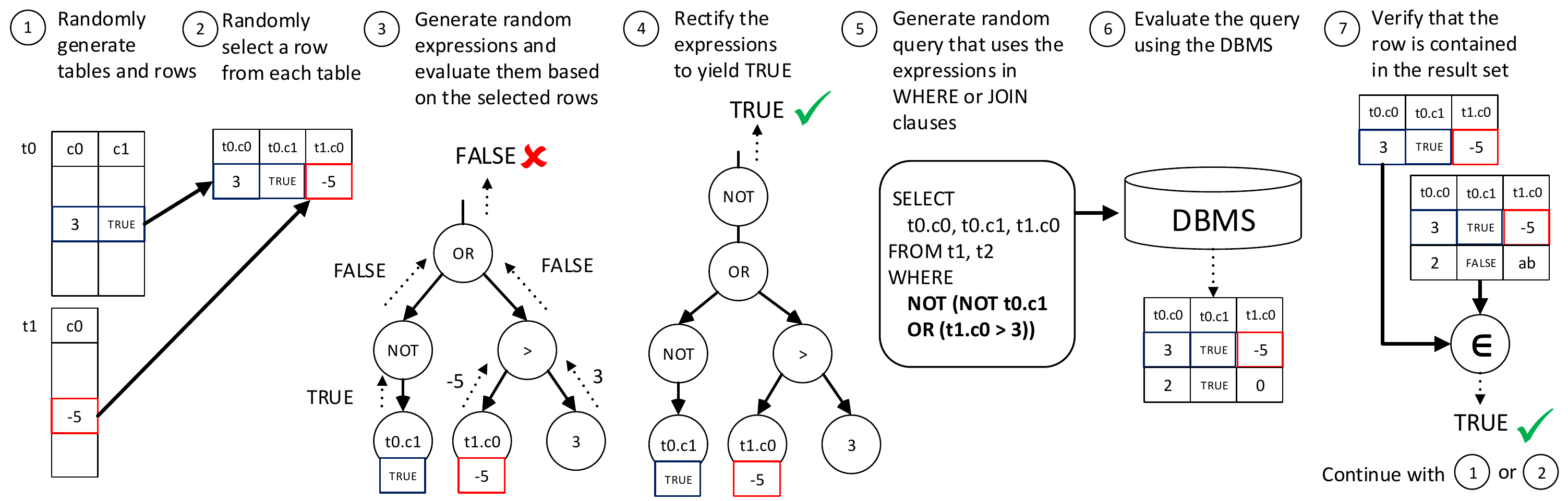}
	\caption{Overview of the Approach implemented in \toolname{}. Dotted lines indicate that a result is generated.}
	\label{fig:overview}
	\reducespace{}

\end{figure*}

\begin{algorithm}
  \caption{The \texttt{generateExpression()} function generates a random AST.}
  \label{alg:astgen}
  \SetKwFunction{FMain}{generateExpression}
  \SetKwProg{Fn}{Function}{:}{}
  \Fn{\FMain{int depth}}{
		$node\_types \leftarrow \{LITERAL$, $COLUMN\}$\\

		\uIf{depth $<$ maxdepth}{
			$node\_types \leftarrow node\_types \cup \{UNARY$, \ldots{}$\}$\\
		}
		$type \leftarrow random(node\_types)$\\
        \Switch{type}{
			\uCase{LITERAL} {
				 \KwRet\ Literal(randomLiteral());
			}
			\uCase{COLUMN} {
				\KwRet\ ColumnValue(randomTable().randomColumn());
			}
			\uCase{UNARY} {
				\KwRet\ UnaryNode(generateExpression(depth+1), UnaryNode.getRandomOperation());
			}
			\uCase{\ldots{}} {
				\ldots{}
			}
		}
 } 
\end{algorithm}

\normalpar{Expression Evaluation}
After building a random expression tree, we must check whether the condition yields \texttt{TRUE} for the \pivot{}.
To this end, every node provides an \texttt{execute()} method that computes the node's result.
Leaf nodes directly return their assigned constant value.
Column nodes are assigned the value that corresponds to their column in the \pivot{}.
For example, in Figure~\ref{fig:overview} step \circled{3}, the leaf node \texttt{t0.c1} returns \texttt{TRUE}, and the constant node \texttt{3} returns an integer \texttt{3}.
Composite nodes compute their result based on the literals returned by their children.
For example, the \texttt{NOT} node returns \texttt{FALSE}, because its child evaluates to \texttt{TRUE} (see Algorithm~\ref{alg:notimpl}).
The node first executes its subexpression, and then casts the result to a boolean; if the result is a boolean value, the value is negated; otherwise \texttt{NULL} is returned.
Note that our implementation is simpler than AST interpreters for programming languages~\cite{Wurthinger2013}, since all nodes operate on literal values (\ie{}, they do not need to consider mutable storage).
Since the bottleneck of our approach is the DBMS evaluating the queries rather than \toolname{}, all operations are implemented naively and do not perform any optimizations.
Some operations require moderate implementation effort nevertheless; for example, the implementation of the \texttt{LIKE} regular expression operator has over 50 LOC in SQLancer.
\parspaceskip{}

\begin{algorithm}
  \caption{The execute() implementation of a \texttt{NOT} node.}
  \label{alg:notimpl}
  \SetKwFunction{FMain}{NotNode::execute}
  \SetKwProg{Fn}{Method}{:}{}
  \Fn{\FMain{}}{%
		$value \leftarrow child.execute()$\\
        \Switch{asBoolean(value)}{
			\uCase{TRUE} {
				$result \leftarrow$ $FALSE$
			}
			\uCase{FALSE} {
				$result \leftarrow$ $TRUE$
			}
			\uCase{NULL} {
				$result \leftarrow$ $NULL$
				}
		}
       \KwRet\ result;
  }
\end{algorithm}

\normalpar{Expression Rectification}
After generating random expressions, step \circled{4} ensures that they evaluate to \texttt{TRUE}.
SQL is based on a three-valued logic.
Thus, when evaluated in a boolean context, an expression either yields \texttt{TRUE}, \texttt{FALSE}, or \texttt{NULL}.
To rectify the expression to yield \texttt{TRUE}, we use Algorithm~\ref{alg:rect}.
For example, in Figure~\ref{fig:overview} step \circled{4}, we modify the expression by adding a preceding \texttt{NOT}, so that the expression evaluates to \texttt{TRUE}.
Note that our approach works also for other logic systems (\eg{}, four-valued logic), by adjusting this step.
\parspaceskip{}

\begin{algorithm}
  \caption{The expression rectification step applied to a randomly-generated expression.}
  \label{alg:rect}
  \SetKwFunction{FMain}{rectifyCondition}
  \SetKwProg{Fn}{Function}{:}{}
  \Fn{\FMain{randexpr}}{
        \Switch{randexpr.execute()}{
			\uCase{TRUE} {
				$result \leftarrow$ $randexpr$
			}
			\uCase{FALSE} {
				$result \leftarrow NOT$ $randexpr$
			}
			\uCase{NULL} {
				$result \leftarrow randexpr$ $ISNULL$
			}
        
        }
        \KwRet\ result;
  }
\end{algorithm}

\reducespace{}


\normalpar{Query generation}
In step \circled{5}, we generate targeted queries that fetch the \pivot{}.
The expressions evaluating to \texttt{TRUE} are used in \texttt{WHERE} clauses, which restrict which rows a query fetches, and in \texttt{JOIN} clauses, which are used to join tables.
Note that \texttt{SELECT} statements typically provide various keywords to control the query's behavior, for example, all DBMS provide a keyword to fetch only distinct values.
We randomly select appropriate keywords when generating these queries.
Note that our approach can also be used to partially test aggregate functions, which compute values over multiple rows, when only a single row is present in a table.
\parspaceskip{}

\normalpar{Checking containment}
After using the DBMS to evaluate the query in step \circled{6}, checking whether the \pivot{} is part of the result set is the last step of our approach.
While the checking routine could have been implemented in \toolname{}, we instead construct the query so that it checks for containment, effectively combining steps \circled{6} and \circled{7}.
Each DBMS provides various operators to check for containment, such as the \texttt{IN} and \texttt{INTERSECT} operators. 
For example, for checking containment in Figure~\ref{fig:overview} step \circled{7}, we can check whether the row \texttt{(3, TRUE, -5)} is contained in the result set using the query \texttt{SELECT (3, TRUE, -5) INTERSECT SELECT t0.c0, t0.c1, t1.c0 FROM t1, t2 WHERE NOT(NOT(t0.c1 OR (t1.c0 > 3)))} in \sqlite{}, which returns a row if the \pivot{} is contained.
\parspaceskip{}

\subsection{Random State Generation}
\label{sec:dbge}
\sloppy{}
In step \circled{1}, we generate a random database state.
We use the \texttt{CREATE TABLE} statement to create tables, and \texttt{INSERT} to insert data rows.
Furthermore, by generating additional DDL and DML statements, we can explore a larger space of databases, some of which exposed DBMS bugs.
For example, we implemented \texttt{UPDATE}, \texttt{DELETE}, \texttt{ALTER TABLE}, and \texttt{CREATE INDEX} commands for all databases, as well as DBMS-specific run-time options.
A number of commands that we implemented were unique to the respective DBMS.
Statements unique to MySQL were \texttt{REPAIR TABLE} and \texttt{CHECK TABLE}.
The statements \texttt{DISCARD} and \texttt{CREATE STATISTICS} were unique to \postgres{}.
\parspaceskip{}

\normalpar{Error handling}
We attempt to generate statements that are correct both syntactically and semantically.
However, generating semantically correct statements is sometimes impractical.
For example, an \texttt{INSERT} might fail when a value already present in an \texttt{UNIQUE} column is inserted again; preventing such an error would require scanning every row in the respective table.
Rather than checking for such cases, which would involve additional implementation effort and a run-time performance cost, we defined a list of error messages that we might expect when executing the respective statement.
Often, we associated an error message to a statement depending on presence or absence of specific keywords; for example, an \texttt{INSERT OR IGNORE} is expected to ignore many error messages that would appear without the \texttt{OR IGNORE}.
If the DBMS returns an expected error, it is ignored.
However, we found a number of cases where an error message was unexpected.
For example, in \sqlite{} a \emph{malformed database disk image} error message is always unexpected, since it indicates the corruption of the database.
Based on this observation, we propose a secondary error oracle, which we termed \emph{\errororacle{}}, and which detects a bug when an unexpected error is caused.

\subsection{Important Implementation Details}
\label{sec:impl}

This section explains implementation decisions, which we consider significant for the outcome of our study.
\parspaceskip{}

\normalpar{Performance}
We optimized \toolname{} to take advantage of the underlying hardware.
We parallelized the system by running each thread on a distinct database, which also resulted in bugs connected to race conditions being found.
To fully utilize each CPU, we decreased the probability of SQL statements being generated that cause low CPU utilization (such as \texttt{VACUUM} in \postgres{}).
Typically, \toolname{} generates 5,0000 to 20,000 statements per second, depending on the DBMS under test.
We implemented the system in Java. 
However, any other programming language would have been equally well suited, as the performance bottleneck was the DBMS executing the queries.
\parspaceskip{}

\normalpar{Number of rows}
We found most bugs by restricting the number of rows inserted to a low value (10--30 rows). A higher number would have caused queries to time out when tables are joined without a restrictive join clause. For example, in a query \texttt{SELECT * FROM t0, t1, t2}, the largest result set for 100 rows in each table would already be $|t0|*|t1|*|t2|=1,000,000$, significantly lowering the query throughput.
\parspaceskip{}

\normalpar{Database state} For the generation of many SQL statements, knowledge of the database schema or other database state is required; for example, to insert data, \toolname{} must determine the name of a table and its columns.
We query such state dynamically from the DBMS, rather than tracking or computing it ourselves, which would require additional implementation effort.
For example, to query the name of the tables, both MySQL and \postgres{} provide an information table \texttt{information\_schema.tables} and \sqlite{} a table \texttt{sqlite\_master}.
\parspaceskip{}

\normalpar{Expressions on columns}
While our initial implementation only checked the containment of the pivot row, we subsequently extended it to also check whether expressions on columns are evaluated correctly.
To achieve this, we allow the randomly-generated query to not only refer to a column, but also to randomly-generated expressions that are potentially based on column references.
Thus, rather than checking whether the pivot row is contained in the result set, we evaluate the expressions based on the pivot row to check whether the expression results are contained in the result set.
\parspaceskip{}


\section{Evaluation}
We evaluated whether the proposed approach is effective in finding bugs in DBMS.
We expected it to detect logic bugs, which cannot be found by fuzzers, rather than crash bugs.
This section overviews the experimental setup, bugs found, and characterizes the SQL statements used to trigger the bugs.
We then present a DBMS-specific bug overview, where we present interesting bugs and bug trends.
To put these findings into context, we measured the size of \toolname{}'s components and the coverage it reaches on the tested DBMS.

\subsection{Experimental Setup}
To test the effectiveness of our approach, we implemented \toolname{} and tested \sqlite{}, MySQL, and \postgres{} in a period of about three months.
Typically, we enhanced \toolname{} to test a new operator or DBMS feature, let the tool run for several seconds up to a day, and then report any new bugs found during this process.
Where possible, we waited for bug fixes before continuing testing and implementing new features.
\parspaceskip{}

\normalpar{Baseline}
Note that there is no applicable baseline to which we could compare our work.
RAGS~\cite{slutz1998massive}, which was proposed more than 20 years ago, would be the closest related work, but is not publicly available and might be outdated.
Due to the small common SQL core, we would expect that RAGS could not find most of the bugs that we found.
Khalek et al. worked on automating testing DBMS using constraint solving~\cite{khalek2008,khalek2010querygen}, with which they found a previously unknown bug.
Also their tool is not available publicly.
SQLsmith~\cite{sqlsmith}, AFL~\cite{afl} as well as other random query generators and fuzzers~\cite{Poess2004} only detect crash bugs in DBMS.
Thus, the only potential overlap between these tools and SQLancer would be the crash bugs that we found, which are not the focus of this work.
\parspaceskip{}

\normalpar{DBMS versions}
For all DBMS, we started testing the latest release version, which was \sqlite{} 3.28, MySQL 8.0.16, and \postgres{} 11.4.
For \sqlite{}, we switched to the latest trunk version (\ie{}, the latest non-release version of the source code) after the first bugs were fixed.
For MySQL, we also tested version 8.0.17 after it was released.
For \postgres{}, we switched to the latest beta version (\postgres{} Beta 2) after opening duplicate bug reports.
Eventually, we continued to test the latest trunk version.
\parspaceskip{}

\normalpar{Bug reporting}
In the \sqlite{} bug tracker, bugs can only be created by \sqlite{} developers, so initially we reported bugs on the public mailing list.
Later, we were offered access to the bug tracker and proceeded to report bugs there.
For MySQL, we reported non-security MySQL bugs on the public bug tracker.
For \postgres{}, we reported non-security bugs on the public mailing list, since \postgres{} lacks a public bug tracker.
We reported crash bugs privately, because we were unsure whether they were security relevant; however, we did not investigate any of the bugs in terms of their security impact, as the focus of this work were logic bugs. 
The test cases that we used in our bug reports were reduced ones; \toolname{} automatically deletes SQL statements that are unnecessary to reproduce a bug, and we also manually shortened them were possible.
Note that all bug reports are documented at \bugurl{}.

\subsection{Bug Reports Overview}

\begin{table}
\tablesize{}
\center
\caption{Total number of reported bugs and their status}
\label{tbl:nrfoundbugs}
\begin{tabular}{l r r r r r}  
\toprule
 &  &  & \multicolumn{2}{c}{Closed} \\
\cmidrule(r){4-5}
DBMS & Fixed & Verified & Intended & Duplicate \\
\midrule
\sqlite{} & \sqlitefixedInDocsOrCode & \sqliteverified{} & \sqliteclosednotabug{} & \sqliteclosedduplicate{}\\  
MySQL &  \mysqlfixedInDocsOrCode & \mysqlverified{} & \mysqlclosednotabug{} & \mysqlclosedduplicate{} \\  
\postgres{} & \postgresfixedInDocsOrCode & \postgresverified{} & \postgresclosednotabug{} & \postgresclosedduplicate{} \\ 
\bottomrule
\end{tabular}
\reducespace{}
\end{table}

Table~\ref{tbl:nrfoundbugs} shows the number of bugs that we reported (\nrReported{} overall).
We considered \nrTruePositives{} bugs as true bugs, because they resulted in code fixes (\fixed{} reports), documentation fixes (\fixedindocumentation{} reports), or were confirmed by the developers (\verified{} reports).
We opened \nrFalsePositives{} bug reports that we classified as false bugs, because behavior exhibited in the bug reports was considered to work as intended (\closednotabug{} reports) or because bugs that we reported were considered to be duplicates (\closedduplicate{} reports, \eg{}, because a bug had already been fixed on the latest non-release version).
\parspaceskip{}

\normalpar{Severity levels}
Only for \sqlite{}, bugs were assigned a severity level by the DBMS developers.
\sqliteNrCriticalBugs{} bugs were classified as \emph{Critical}, \sqliteNrSevereBugs{} bugs as \emph{Severe}, and \sqliteNrImportantBugs{} as \emph{Important}.
For \sqliteNrNoneBugs{} bugs, we reported them on the mailing list and no entry in the bug tracker was created.
The other bug reports were assigned low severity levels such as \emph{Minor}.
While the severity level was not set consistently, this still provides evidence that we found many critical bugs.
For the other DBMS, we lack data on how severe the bugs were.
\parspaceskip{}

\normalpar{\sqlite{} bug handling}
For \sqlite{}, the main developers reacted to most of our bug reports shortly after reporting them, and fixed issues typically within a day, which is why is why no \emph{verified} and \emph{open} bugs are listed.
The developers' quick responses was a significant factor for the high number of bugs that we reported for \sqlite{}, which led us to focus our testing efforts on this DBMS.
For \sqlite{}, we also tested \texttt{VIEWS}, non-default \texttt{COLLATE}s (which define how strings are compared), floating-point support, and aggregate functions, which we omitted for the other DBMS.
\parspaceskip{}

\normalpar{MySQL bug handling}
For MySQL, bug reports were typically verified within a day by a tester.
This tester also evaluated whether the bug could be reproduced on other MySQL versions than the one we specified.
MySQL's development is not open for the general public.
Although we tried to establish contact with MySQL developers, we could not obtain any information that went beyond what is visible on the public bug tracker.
Thus, it is likely that some of the verified bug reports will subsequently be considered as duplicates or classified to work as intended.
Furthermore, although MySQL is available as open-source software, only the code for the latest release version is provided, so any bug fixes could be verified only with the subsequent release.
This was a significant factor that restricted us in finding bugs in MySQL; due to the increased effort of verifying whether a newly found bug was already reported, we invested limited effort into testing MySQL.
\parspaceskip{}

\normalpar{\postgres{} bug handling}
For \postgres{}, we received feedback to bug reports within a day, and it typically took multiple days or weeks until a bug was fixed, since possible fixes and patches were discussed intensively on the mailing list.
As we found less bugs for \postgres{} overall, this did not significantly influence our testing efforts.
Note that not all confirmed bugs were fixed.
For example, for one reported bug, a developer decided to ``put this on the back burner until we have some consensus how to proceed on that''; from the discussion, we speculate that the changes needed to address the bug properly were considered too invasive.
\parspaceskip{}


\normalpar{Test oracles}
Table~\ref{tbl:oracles} shows the test oracles that we used to detect the true bugs.
The \containsoracle{} accounts for most of the bugs that we found, which is expected, since our approach mainly builds on this oracle.
Perhaps surprisingly, the \errororacle{} also contributed a large number of bugs.
We believe that this observation could be used when using fuzzers to test DBMS, for example, by checking for specific error messages that indicate database corruptions.
Our approach also detected a number of crash bugs, one of which was considered a security vulnerability in MySQL.
These bugs are somewhat less interesting, since they could also have been found by traditional fuzzers.
\parspaceskip{}

\begin{table}
\tablesize{}
\caption{The oracles and how many bugs they found.}
\center
\label{tbl:oracles}
\begin{tabular}{l r r r}
\toprule{}
DBMS & Contains & Error & SEGFAULT \\
\midrule
\sqlite{} & \sqlitecontains{} & \sqliteerror{} & \sqlitesegfault{}\\
MySQL & \mysqlcontains{} & \mysqlerror{} & \mysqlsegfault{}\\
\postgres{} & \postgrescontains{} & \postgreserror{} & \postgressegfault{}\\
Sum & \sumcontains{} & \sumerror{} & \sumsegfault{} \\
\bottomrule
\end{tabular}
\reducespace{}

\end{table}

\subsection{SQL Statements Overview}

\begin{figure}
	\includegraphics{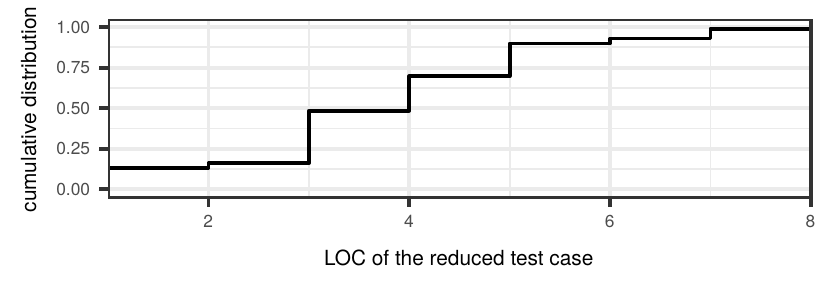}
	\caption{The cumulative distribution of LOC needed to reproduce a bug.}
	\label{fig:cumloc}
	\reducespace{}

\end{figure}

\normalpar{Test case length}
Our test cases typically comprised only a few SQL statements (\avgloc{} LOC on average).
Note that we reduced test cases before reporting them.
Figure~\ref{fig:cumloc} shows the cumulative distribution of the number of statements in a test case to reproduce a bug.
For \nrtestcaseswithsingleline{} test cases, a single line was sufficient.
Such test cases were either \texttt{SELECT} statements that operated on constants, or operations that set DBMS-specific options.
As an example, Listing~\ref{lst:suberr} shows a bug in \sqlite{} where subtracting an integer from a \texttt{TEXT} value resulted in an incorrect result.
As another example, Listing~\ref{lst:option} shows a bug in MySQL where setting an option nondeterministically failed with an error.
The maximum number of statements required to reproduce a bug was \maxloc{}. 
A \postgres{} crash bug that had already been fixed when we reported it required even 27 statements to be reproduced.
\parspaceskip{}

\reducespace{}

\begin{lstlisting}[caption=Bug in \sqlite{} where subtracting an integer from a string produces an incorrect result, label=lst:suberr]
SELECT '' - 2851427734582196970; -- actual: -2851427734582196736, expected: -2851427734582196970
\end{lstlisting}

\reducespace{}

\begin{lstlisting}[caption=Bug in MySQL where setting an option nondeterministically failed with an error,label=lst:option]
SET GLOBAL key_cache_division_limit = 100; -- ERROR 1210 (HY000): Incorrect arguments to SET
\end{lstlisting}


\begin{figure*}
	\includegraphics{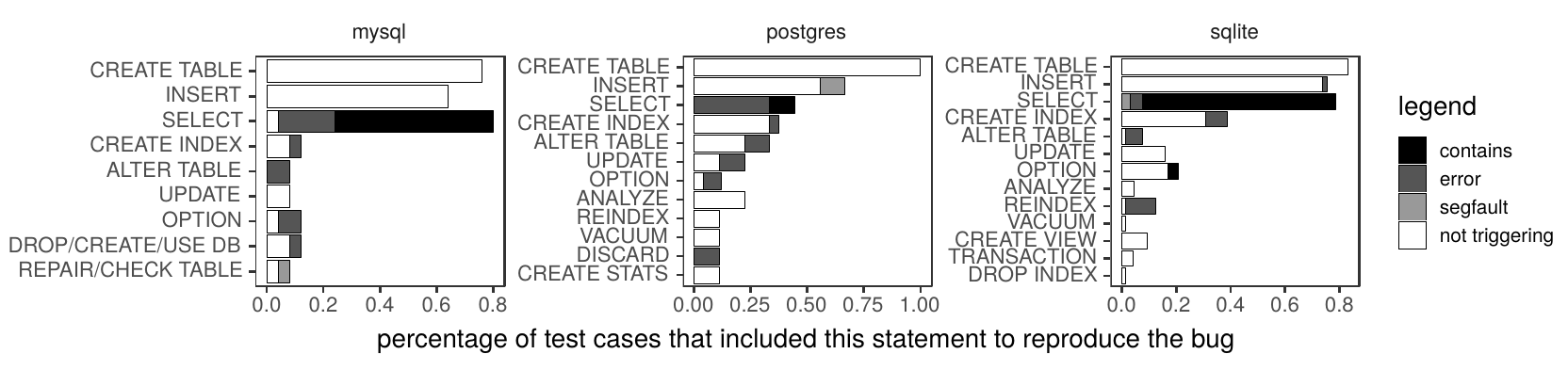}
	\caption{The distribution of the SQL statements used in the bug reports to reproduce the bug. A non-white filling indicates that a statement of the respective category triggered the bug, which was exposed by the test oracle as indicated by the filling (\ie{}, it was the last statement in the bug report).}
	\label{fig:statementdistr}
	\reducespace{}
\end{figure*}

\normalpar{Statement distribution}
Figure~\ref{fig:statementdistr} shows the distribution of statements.
Note that for some bug reports, we had to select the simplest test case among multiple failing ones, which might skew these results.
The \texttt{CREATE TABLE} and \texttt{INSERT statements} are part of most bug reports for all DBMS, which is expected, since only few bugs can be reproduced without manipulating or fetching data from a table.
\percOneCreateTable{}\% of the bug reports included only a single table.
The \texttt{SELECT} statement also ranks highly, since the \containsoracle{} relies on it.
In all DBMS, the \texttt{CREATE INDEX} statements rank highly; especially for \sqlite{}, we reported a number of bugs where creating an index resulted in a malformed database image or in a row not being fetched.
We found that statements that compute or recompute table state were error prone, for example, \texttt{REPAIR TABLE} and \texttt{CHECK TABLE} in MySQL, as well as \texttt{VACUUM} and \texttt{REINDEX} in \sqlite{} and \postgres{}.
DBMS-specific options, such as \texttt{SET} in MySQL and \postgres{}, and \texttt{PRAGMA} in \sqlite{} also resulted in bugs being found.
For \postgres{}, bug reports contained \texttt{ANALYZE}, which gathers statistics to be used by the query planner.
\parspaceskip{}

\normalpar{Column Constraints}
Column constraints, which can be used to restrict the values stored in a column, were often part of test cases.
The most common constraint was \texttt{UNIQUE} (appearing in \percUNIQUE{}\% of the test cases).
Also \texttt{PRIMARY KEY} columns were frequent (\percPRIMARYKEY\%).
Typically, the DBMS enforce \texttt{UNIQUE} and \texttt{PRIMARY KEY} by creating indexes; explicit indexes, created by \texttt{CREATE INDEX} were more common, however (\percINDEX\%).
Other constraints were uncommon, for example, \texttt{FOREIGN KEY}s appeared only in \percFOREIGNKEY{}\% of the bug reports.


\subsection{Bugs in \sqlite{}}

\normalpar{Features}
In \sqlite{}, many bugs resulted from a combination of unique language features.
\sqliteINDEX{} bug reports included indexes, \sqliteCOLLATE{} included COLLATE sequences, and \sqliteWITHOUTROWID{} \texttt{WITHOUT ROWID} tables.
For example, Listing~\ref{lst:sqlitefirstbug} shows a test case for the first bug that we found with our approach, where these features were combined, and where \sqlite{} failed to fetch a row.
The bug was classified as \emph{Severe} and goes back to when \texttt{WITHOUT ROWID} tables were introduced in 2013.
As another example, Listing~\ref{lst:rtrimissue} shows a test case that detected an 11 years old \emph{Important} bug.
This test case uses a \texttt{COLLATE}, and \texttt{WITHOUT ROWID} to expose a bug where \texttt{RTRIM} was implemented incorrectly.
As mentioned initially, \sqlite{} allows storing values of any type in a column, irrespective of its declared type; we discovered \sqliteUNEXPECTEDTYPE{} bugs related to this feature.
For example, Listing~\ref{lst:likeopt} shows a minor bug where an optimization for the \texttt{LIKE} operator was implemented incorrectly when applied to \texttt{INT} values.

\begin{lstlisting}[label=lst:sqlitefirstbug,caption={The first bug that we found with our approach involved a \texttt{COLLATE} index, and a \texttt{WITHOUT ROWID} table.}]
CREATE TABLE t0(c0 TEXT PRIMARY KEY) WITHOUT ROWID;
CREATE INDEX i0 ON t0(c1 COLLATE NOCASE);
INSERT INTO t0(c0) VALUES ('A');
INSERT INTO t0(c0) VALUES ('a');
SELECT * FROM t0; -- unexpected: only one row is fetched
\end{lstlisting}

\reducespace{}

\begin{lstlisting}[label=lst:rtrimissue,caption=This test case demonstrates an 11 years old \texttt{RTRIM} bug.]
CREATE TABLE t0(c0 COLLATE RTRIM, c1 BLOB UNIQUE, PRIMARY KEY (c0, c1)) WITHOUT ROWID;
INSERT INTO t0 VALUES (123, 3), (' ', 1),
	('	', 2), ('', 4);
SELECT * FROM t0 WHERE c1 = 1; -- expected: ' ', 1, actual: no row is fetched
\end{lstlisting}

\normalpar{Incorrect optimizations}
A number of bugs could be traced back to incorrect optimizations.
For example, Listing~\ref{lst:skipscan} shows a test case that demonstrated that the skip-scan optimization, where an index is used even if its columns are not part of the \texttt{WHERE} clause, was implemented incorrectly for \texttt{DISTINCT} queries.
The bug was classified as \texttt{Severe}.
As another example, we found \sqliteLIKEOPTIMIZATION{} minor bugs in the implementation of an optimization for \texttt{LIKE} and no-text affinity.
Listing~\ref{lst:likeopt} demonstrates an example where an exact string match incorrectly yielded \texttt{FALSE}.

\begin{lstlisting}[caption=\sqlite{}'s skip-scan optimization was implemented incorrectly for \texttt{DISTINCT},label=lst:skipscan]
CREATE TABLE t1 (c1 , c2, c3, c4,
	PRIMARY KEY (c4, c3));
INSERT INTO t1(c3) VALUES (0), (0), (0), (0), (0), (0), (0), (0), (0), (0), (NULL), (1), (0);
UPDATE t1 SET c2 = 0;
INSERT INTO t1(c1) VALUES (0), (0), (NULL),
	(0), (0);
ANALYZE t1;
UPDATE t1 SET c3 = 1;
SELECT DISTINCT * FROM t1 WHERE t1.c3 = 1; -- expected: |0|1|, 0||1|, ||1|, actual: |0|1|
\end{lstlisting}

\reducespace{}

\begin{lstlisting}[label=lst:likeopt,caption={We discovered \sqliteLIKEOPTIMIZATION{} bugs in a \texttt{LIKE} optimization, one demonstrated by this test case.}]
CREATE TABLE t0(c0 INT UNIQUE COLLATE NOCASE);
INSERT INTO t0(c0) VALUES ('./');
SELECT * FROM t0 WHERE t0.c0 LIKE './'; -- unexpected: fetches no rows
\end{lstlisting}

\normalpar{Language Deficiencies}
\sqliteLANGUAGEDEFICIENCY{} bug reports uncovered issues in \sqlite{}'s SQL dialect. 
The test case shown in Listing~\ref{lst:sqllanguagechange} caused the SQL developers to disallow strings in double quotes when creating indexes.
\sqlite{} allows both single quotes and double quotes to be used to denote strings; depending on the context, either can refer to a column name.
In the example, after the \texttt{RENAME} operation, it is ambiguous whether the index refers to a string or column, and the \texttt{SELECT} fetches C3 as a value for the column c3, which is incorrect in either case.
As another example, Listing~\ref{lst:malformedschema} shows a test case that causes the database schema to disagree with the database content, because the behavior of the \texttt{LIKE} operator could be controlled by a run-time flag.
The developers stated that this was ``a defect in the design of SQLite, not a defect in the implementation''.
Seven options to tackle this were outlined, but eventually the issue was merely documented and a new compile-time option to disable the \texttt{PRAGMA} was added.
\reducespace{}

\begin{lstlisting}[caption=This bug report caused the \sqlite{} developers to disallow double quotes in indexes.,label=lst:sqllanguagechange]
CREATE TABLE t0(c1, c2);
INSERT INTO t0(c1, c2) VALUES  ('a', 1);
CREATE INDEX i0 ON t0("C3");
ALTER TABLE t0 RENAME COLUMN c1 TO c3;
SELECT DISTINCT * FROM t0; -- fetches C3|1 rather than a|1
\end{lstlisting}
\reducespace{}

\begin{lstlisting}[label=lst:malformedschema,caption={The \texttt{PRAGMA case\_sensitive\_like} can cause mismatches between the database schema and database content.}]
CREATE TABLE test (c0);
CREATE INDEX index_0 ON test(c0 LIKE '');
PRAGMA case_sensitive_like=false;
VACUUM;
SELECT * from test; -- Error: malformed database schema (index_0) - non-deterministic functions prohibited in index expressions
\end{lstlisting}

\normalpar{Bugs found with the \errororacle{}}
We discovered \sqliteerror{} bugs using the \errororacle{}, the most severe ones being those that corrupted the database, which happened in \sqliteCORRUPTEDDB{} cases, as indicated by \emph{malformed database schema} errors.
For example, Listing~\ref{lst:malformedreal} shows a test case where manipulating values in a \texttt{REAL PRIMARY KEY} column resulted in a corrupted database.
The bug was introduced in 2015, and went undetected until we reported it in 2019; it was assigned a \emph{Severe} severity level.
Another common trigger was \texttt{REINDEX}, which detected violated constraints, resulting in errors such as \texttt{UNIQUE constraint failed}, with which we found \sqliteREINDEXCONSTRAINTFAILEDwithFP{} bugs.

\reducespace{}

\begin{lstlisting}[label=lst:malformedreal,caption={We found \sqliteCORRUPTEDDB{} malformed database errors using the error oracle, such as this one.}]
CREATE TABLE t1 (c0, c1 REAL PRIMARY KEY);
INSERT INTO t1(c0, c1) VALUES (TRUE, 9223372036854775807), (TRUE, 0);
UPDATE t1 SET c0 = NULL;
UPDATE OR REPLACE t1 SET c1 = 1;
SELECT DISTINCT * FROM t1 WHERE (t1.c0 IS NULL); -- database disk image is malformed
\end{lstlisting}

\subsection{Bugs in MySQL}

\normalpar{Engine-specific bugs}
MySQL provides various engines that can be assigned to tables, a feature that is not provided by the other DBMS we tested.
The default engine is the \texttt{InnoDB} engine; an example for an alternative engine is the \texttt{MEMORY} engine, which stores data in-memory rather than on disk.
We found \mySQLNrEngineBugs{} bugs that were triggered only using such alternative engines.
Listing~\ref{lst:enginespecific} shows a test case where a row is not fetched using the \texttt{MEMORY} engine.
\reducespace{}

\begin{lstlisting}[label=lst:enginespecific,caption={We found \mySQLNrEngineBugs{} bugs using non-default engines}]
CREATE TABLE t0(c0 INT);
CREATE TABLE t1(c0 INT) ENGINE = MEMORY;
INSERT INTO t0(c0) VALUES(0);
INSERT INTO t1(c0) VALUES(-1);
SELECT * FROM t0, t1 WHERE (CAST(t1.c0 AS UNSIGNED)) > (IFNULL("u", t0.c0)); -- expected: row is fetched, actual: no row is fetched
\end{lstlisting}

\normalpar{Unsigned integer bugs}
Unlike the other DBMS, MySQL provides unsigned integer data types.
We found \mySQLNrUnsignedBugs{} bugs related to this feature.
For example, also Listing~\ref{lst:enginespecific} relies on a cast to \texttt{UNSIGNED}.
\parspaceskip{}

\normalpar{Value range bugs}
We found a number of bugs where queries were handled incorrectly depending on the magnitude of an integer or floating-point number.
For example, Listing~\ref{lst:mysql} shows a bug where the MySQL-specific \texttt{<=>} inequality operator, which yields a boolean value even when an argument is \texttt{NULL}, yielded \texttt{FALSE} when the column value was compared with a constant that was greater than what the column's type can represent.
Before the release of MySQL version 8.0.17, we were informed that this would be fixed for 8.0.18.
As another example, we found that small double values (\eg{}, \texttt{0.5}) stored in \texttt{TEXT} columns incorrectly evaluated to \texttt{FALSE} when used in a boolean context.
One such bug was fixed for version 8.0.17.
\parspaceskip{}

\reducespace{}

\begin{lstlisting}[caption=Custom comparison operator results in incorrect result,label=lst:mysql]
CREATE TABLE t0(c0 TINYINT);
INSERT INTO t0(c0) VALUES(NULL);
SELECT * FROM t0 WHERE NOT(t0.c0 <=> 2035382037);
\end{lstlisting}



\normalpar{Duplicate bugs}
In one case, which we considered as a duplicate, a bug seems to have been fixed already in a version not released to the public (see Listing~\ref{lst:mysqldoubleneg}).
Applying \texttt{NOT} to a non-zero integer value should yield 0, and negating 0 should yield 1.
However, it seems that MySQL optimized away the double negation, which would be correct for boolean values, but not for other data types, resulting in the row not being fetched.
We believe that the implicit conversions provided by MySQL (and also \sqlite{}) is one of the reasons that we found more bugs in these DBMS than in \postgres{}.
\parspaceskip{}

\reducespace{}

\begin{lstlisting}[caption=Double negation bug,label=lst:mysqldoubleneg]
CREATE TABLE t0(c0 INT);
INSERT INTO t0(c0) VALUES(1);
SELECT * FROM t0 WHERE 123 != (NOT (NOT 123)); -- expected: row is fetched, actual: row is not fetched
\end{lstlisting}

\normalpar{Segfault}
We found one \texttt{SEGFAULT} bug in MySQL, which was triggered when executing a sequence of SQL statements using multiple threads (see Listing~\ref{lst:mysqlcve}).
To trigger this bug, the \texttt{CHECK TABLE} statement had to be used, which is unique to MySQL.
After reporting this error to Oracle, it received a CVE entry (CVE-2019-2879) and was classified as a medium security vulnerability (CVSS v3.0 Base Score 4.9).

\parspaceskip{}

\begin{lstlisting}[caption=SEGFAULT bug in MySQL,label=lst:mysqlcve]
CREATE TABLE t0(c0 INT);
CREATE INDEX i0 ON t0((t0.c0 || 1));
INSERT INTO t0(c0) VALUES(1);
CHECK TABLE t0 FOR UPGRADE;
\end{lstlisting}

\subsection{Bugs in \postgres{}}

In \postgres{}, using our \containsoracle{}, we found only \postgrescontains{} bug that was fixed.
The bug was related to table inheritance, a feature that only \postgres{} provides (see Listing~\ref{lst:tableinheritance}).
Table \texttt{t1} inherits from \texttt{t0}, and \postgres{} merges the \texttt{c0} column in both tables.
As described in the \postgres{} documentation, t1 does not respect the PRIMARY key restriction of t0.
This was not considered when implementing the \texttt{GROUP BY} clause, which caused \postgres{} to omit one row in its result set.
\parspaceskip{}

\reducespace{}

\begin{lstlisting}[caption=Table inheritance bug in \postgres{}, label=lst:tableinheritance]
CREATE TABLE t0(c0 INT PRIMARY KEY, c1 INT);
CREATE TABLE t1(c0 INT) INHERITS (t0);
INSERT INTO t0(c0, c1) VALUES(0, 0);
INSERT INTO t1(c0, c1) VALUES(0, 1);
SELECT c0, c1 FROM t0 GROUP BY c0, c1; -- expected: 0|0 and 0|1, actual: 0|0
\end{lstlisting}

We found the other \postgreserror{} bugs using the \errororacle{}.
For example, Listing~\ref{lst:negativebitmap} shows a test case where a \texttt{WHERE} condition triggered a bug resulting in an error \emph{negative bitmapset member not allowed}.
After we reported the bug, on the same day, another independent bug report was created based on a finding of \sqlsmith{}, which caused \postgres{} to crash based on the same underlying bug.
This provides further evidence that DBMS are being extensively tested and fuzzed.
Note that we also found and reported two structurally different crash bugs that exposed this issue, which we later classified as duplicates.
\parspaceskip{}

\reducespace{}

\begin{lstlisting}[caption=Negative bitmapset member error in \postgres{},label=lst:negativebitmap]
CREATE TABLE t0(c0 serial, c1 boolean);
CREATE STATISTICS s1 ON c0, c1 FROM t0;
INSERT INTO t0(c1) VALUES(TRUE);
ANALYZE;
CREATE INDEX i0 ON t0(c0, (t0.c1 AND t0.c1));
SELECT * FROM (SELECT t0.c0 FROM t0 WHERE (((t0.c1) AND (t0.c1)) OR FALSE) IS TRUE) as result WHERE result.c0 IS NULL; -- unexpected: ERROR:  negative bitmapset member not allowed
\end{lstlisting}

\normalpar{Multithreaded bugs}
\postgresMULTITHREADEDwithFP{} reported bugs (including closed/duplicate ones) could only be reproduced when running multiple threads.
For example, Listing~\ref{lst:unexpectednull} shows a bug that was triggered only when another thread opened a transaction, holding a snapshot with the \texttt{NULL} value.
In order to reproduce such bugs, we had to record traces of all executing threads.
In some cases, reducing or reproducing a bug was impractical; for example, we encountered a memory leak that could be observed only after running \postgres{} for a long time.
\parspaceskip{}

\reducespace{}

\begin{lstlisting}[caption=Unexpected null value bug in \postgres{},label=lst:unexpectednull]
CREATE TABLE t0(c0 TEXT);
INSERT INTO t0(c0) VALUES('b'), ('a');
ANALYZE;
INSERT INTO t0(c0) VALUES (NULL);
UPDATE t0 SET c0 = 'a';
CREATE INDEX i0 ON t0(c0);
SELECT * FROM t0 WHERE 'baaaaaaaaaaaaaaaaaaaaaaaaaaaaaaaaaaaaaaaaaaa' > t0.c0; -- unexpected: ERROR: found unexpected null value in index "i0"
\end{lstlisting}


\normalpar{False positives} 
\postgresclosednotabug{} bug reports were closed since they were not considered to be bugs, which is a larger number than for the other DBMS.
We believe that this is partly due to the \postgres{} developer's pragmatic approach towards handling and fixing bugs.
For example, we found that running \texttt{VACUUM FULL} on distinct databases can cause deadlocks to occur.
We created a test case that runs 32 threads to reproduce the deadlock quickly.
Responses on the mailing list concluded that routine use of \texttt{VACUUM FULL} should be avoided, even more so running 32 threads at once.
Listing~\ref{lst:vacuum} shows another issue that we reported that requires another thread holding a snapshot of the value \texttt{2147483647} in the table.
We found that the \texttt{VACUUM} fails with an error caused by an integer overflow, which was surprising to us, since we did not expect \texttt{VACUUM} to fail.
As explained by a \postgres{} developer, an optimization caused the index to not be built for that row, so that the issue only surfaced when using the \texttt{VACUUM}.
Although this was admitted to be somewhat surprising, addressing this would have had other downsides that the developers wanted to avoid.
\parspaceskip{}

\reducespace{}
\begin{lstlisting}[caption=VACUUM issue in \postgres{}, label=lst:vacuum]
CREATE TABLE t1(c0 int);
INSERT INTO t1(c0) VALUES(2147483647);
UPDATE t1 SET c0 = 0;
CREATE INDEX i0 ON t1((1 + t1.c0));
VACUUM FULL; -- unexpected: ERROR: integer out of range
\end{lstlisting}

\normalpar{Duplicates}
\postgresclosedduplicate{} bugs were classified as duplicates.
We reported two structurally-different crash bugs for the latest beta release (12beta2), which were due to the same underlying bug and had already been fixed in the latest trunk version.
The other two were crash bugs and duplicates of the bug triggered by Listing~\ref{lst:negativebitmap}.
As previously explained, a crash bug on the same issue was reported independently.
We believe that these duplicates show that our approach can find bugs that are relevant in practice, and are reported by other users or developers.
\parspaceskip{}

\subsection{Implementation Size and Coverage}

\normalpar{Implementation effort}
It is difficult to quantify the effort that we invested for implementing support for each DBMS, since, for example, we got more efficient in implementing support over time.
The LOC of code of the individual testing components (see Table~\ref{tbl:locimpl}) reflects our estimates that we invested the most effort to test \sqlite{}, then \postgres{}, and then MySQL.
The code part shared by the components is rather small (\numprint{\locCommon{}} LOC), which provides some evidence for the different SQL dialects that they support.
We believe that the implementation effort for \toolname{} is small when compared to the size of the tested DBMS.
Note that the LOC in this table were derived after compiling the respective DBMS using default configurations, and thus include only those lines reachable in the binary.
Thus, they are significantly smaller than the ones we derived statically for the entire repositories in Table~\ref{tbl:dbmsoverview}.
\parspaceskip{}

\begin{table}
\tablesize{}
\caption{The size of \toolname{}'s components specific and common to the tested databases.}
\label{tbl:locimpl}
\begin{tabular}{l r r r r r}
\toprule{}
& \multicolumn{2}{c}{LOC} & \multicolumn{3}{c}{Coverage} \\
\cmidrule(r){2-4}\cmidrule(r){5-6}
DBMS & \toolname{} & DBMS  & Ratio & Line & Branch\\
\midrule
\sqlite{} & \numprint{\locSqlite{}} & \numprint{\locDBMSSqlite} & \implPercSqlite{}\% & 43.0\% & 38.4\%\\
MySQL & \numprint{\locMysql{}} & \numprint{\locDBMSMySQL}  & \implPercMysql\% & 24.4\% & 13.0\% \\
\postgres{} & \numprint{\locPostgres{}} & \numprint{\locDBMSPostgres} & \implPercPostgres\%  & 23.7\% & 16.6\%\\
\bottomrule
\end{tabular}
\reducespace{}
\end{table}

\normalpar{Coverage}
To obtain an estimate on how much code of the DBMS we tested, we instrumented each DBMS and ran \toolname{} for 24 hours on it (see Table~\ref{tbl:locimpl}).
The coverage appears to be low (less than 50\% for all DBMS); however, this is expected, because we were only concerned about testing data-centric SQL statements.
MySQL and \postgres{} provide features such as user management, replication, and database maintenance functionalities, which we did not test.
Furthermore, all DBMS provide consoles to interact with the DBMS and programming APIs.
Furthermore, \toolname{} still lacks support for many common DBMS features.
For example, we currently do not test many data types, language elements such transaction savepoints, many DBMS-specific functions, configuration options that can only be set at server start-up time, and operations that might conflict with other threads running on a distinct database.
The coverage for \sqlite{} is the highest, reflecting that we invested most effort in testing it, but also that it provides fewer features in addition to its SQL implementation.
\parspaceskip{}

\section{Discussion}

\normalpar{Number of Bugs and Code Quality}
The number of bugs that we found in the respective DBMS depended on many, difficult-to-quantify factors.
We found most bugs in \sqlite{}.
A significant reason for this is that we focused on this DBMS, because the developers quickly fixed all bugs.
Furthermore, while the SQL dialect supported by \sqlite{} is compact, we perceived it to be the most flexible one, as, for example, column types are not enforced, leading to bugs that were not present in \postgres{}, and to a lesser degree in MySQL.
MySQL's release policy made it difficult to test it efficiently, limiting the number of bugs that we found in this DBMS.
In \postgres{}, we found the least number of bugs, and we believe that a significant reason for this is that the SQL dialect support is strict, and, for example, only performs few implicit conversions.
\parspaceskip{}

\normalpar{Existing test efforts}
All three DBMS are extensively tested.
For example, \sqlite{}, for which we found most bugs, has 662 times as much test code and test scripts than source code~\cite{sqlitetesting}.
The core is tested by three separate test harnesses.
The TCL tests comprise 45K test cases, the TH3 proprietary test harness contains about 1.7 million test instances and provides 100\% branch test coverage and 100\% MC/DC test coverage~\cite{KellyJ2001}, and the SQL Logic Test runs about 7.2 million queries based on over 1 GB of test data.
\sqlite{} uses various fuzzers such as a random query generator called \emph{SQL Fuzz}, a proprietary fuzzer \emph{dbsqlfuzz}, and it is fuzzed by Google's OSS Fuzz project~\cite{ossfuzz}.
Other kinds of tests are also applied, such as crash testing, to demonstrate that the database will not go corrupt on system crashes or power failures.
Considering that \sqlite{} and other DBMS are tested this extensively, we believe that it is surprising that \toolname{} could find any bugs.
\parspaceskip{}

\normalpar{Reception}
The DBMS developers appreciated our work and effort.
For example, for one DBMS, the developers reached out to actively support us in finding new bugs.
As an anecdote, after a bug that we reported was not fixed within 2 weeks, they also contacted us to ask whether we had stopped testing the DBMS; it turned out that the bug report was overlooked, but then quickly fixed, indicating the importance of our work.
For another DBMS, we were told that it is ``not often we get that many true positives from a tool. We do run fuzzers, but it's not common to find that many bugs in such a short time.''
\parspaceskip{}


\normalpar{Relational databases}
Although relational DBMS are the most common form of DBMS, other models also exist, many to which our approach could be applied.
NoSQL DBMS are based on various non-relational, or partly-relational data models~\cite{Cattell2011}.
For example, MongoDB~\cite{mongodb} is a popular, document-oriented DBMS, and thus stores \emph{documents} rather than rows, where each document describes the data (rather than a schema) and holds the data.
Our technique could be applied to such a DBMS by selecting random data in a randomly-selected document and then constructing a query so that the data should be selected.
\parspaceskip{}

\normalpar{Implementation effort}
Since the supported SQL dialects differ vastly between DBMS, we had to implement DBMS-specific components in \toolname{}.
It could be argued that the implementation effort is too high, especially when the full support of a SQL dialect is to be tested, which could arguably be similar to implementing a new DBMS.
Indeed, we could not test complex functions such as \sqlite{}'s \texttt{printf}, which would have required significant implementation effort.
However, we still argue that the implementation effort is reasonably low, and allows testing significant parts of a DBMS.
Most significantly, our approach effectively evaluates only literal expressions, and does not need to consider multiple rows.
This obviates the need of implementing a query planner, which typically is the most complex component of a DBMS~\cite{giakoumakis2008testing}.
Furthermore, the performance of the evaluation engine is insignificant; the performance bottleneck was the DBMS evaluating the queries, rather than \toolname{}.
Thus, we also did not implement any optimizations, which typically require much implementation effort in DBMS~\cite{Graefe:1993}.
Finally, we did not need to consider aspects such as concurrency and multi-user control as well as integrity.

\parspaceskip{}

\normalpar{Checking a single row}
By checking one row at a time, rather than all the rows, our approach is simple to implement.
To compute and evaluate a \texttt{WHERE} condition, only operations on constants need to be performed, based on the \pivot{}.
Nevertheless, our approach is, in principle, mostly as effective as an approach that checks all rows, considering that the same SQL statements can be generated for all rows in a table, albeit requiring multiple steps.
The only obvious conceptual limitation is that we cannot detect logic bugs where a DBMS erroneously fetches duplicate rows.
\parspaceskip{}

\section{Related Work}

\normalpar{Testing of Software Systems}
This paper fits into the stream of testing approaches for important software systems.
Differential testing~\cite{mckeeman1998differential} is a technique that compares the results obtained by multiple systems that implement a common language; if results deviate, one or multiple of the systems are likely to have a bug.
It has been used as a basis for many approaches, for example, to test C/C++ compilers~\cite{Yang2011,Zhang2017}, symbolic execution engines~\cite{Kapus2017}, and PDF readers~\cite{Kuchta2018}.
Metamorphic testing~\cite{Chen1998}, where the program is transformed so that the same result as for the original program is expected, has been applied to various systems; for example, \emph{equivalence modulo inputs} is a metamorphic-testing-based approach that has been used to find over one thousand bugs in widely-used compilers~\cite{Le2014}.
As another example, metamorphic testing has been successfully applied to test graphic shader compilers~\cite{Donaldson2017}.
We present \emph{\approach{}} as a novel approach to testing DBMS, which solves the \emph{oracle problem} in a novel way, namely by checking whether a DBMS works correctly for a specific query and row.
We believe that our approach can also be extended to test other software systems that have an internal state, of which a single instance can be selected.
\parspaceskip{}

\normalpar{Differential Testing of DBMS}
Slutz proposed an approach \emph{RAGS} for finding bugs in DBMS based on differential testing~\cite{slutz1998massive}.
In \emph{RAGS}, queries are automatically generated and evaluated by multiple DBMS.
If the results are inconsistent, a bug has been found.
As stated in his paper, the approach was very effective, but is applicable to only a small set of common SQL statements.
In particular, the differences in NULL handling, character handling, and numeric type coercions were mentioned as problematic.
Our approach can detect bugs also in SQL statements unique to a DBMS, but requires separate implementations for each DBMS.
\parspaceskip{}

\normalpar{Database Fuzzing}
\sqlsmith{} is a popular tool that randomly generates SQL queries to test various DBMS~\cite{sqlsmith}.
\sqlsmith{} has been highly successful and has found over 100 bugs in popular DBMS such as \postgres{}, \sqlite{} and MonetDB since 2015.
However, it cannot find logic bugs found by our approach.
Similarly, general-purpose fuzzers such as AFL~\cite{afl} are routinely applied to DBMS, and have found many bugs, but also cannot detect logic bugs.
\parspaceskip{}

\normalpar{Queries satisfying constraints}
A number of approaches improved upon random query generation by generating queries that satisfy certain constraints, such as cardinalities or coverage characteristics.
The problem of generating a query, whose subexpressions must satisfy certain constraints, has been extensively studied~\cite{Bruno2006,Mishra2008}; since this problem is complex, it is typically tackled by an approximate algorithm~\cite{Bruno2006,Mishra2008}.
An alternative approach was proposed by Bati et al. where queries are selected and mutated based on whether they increase the coverage of rarely executed code paths~\cite{bati2007}, increasing the coverage of the DBMS component under test.
Rather than improved query generation, Lo et al. proposed an approach where a database is generated based on specific requirements on test queries~\cite{Lo2010framework}.
While these approaches improve the query and database generation, they do not help in automatically finding errors, since they do not propose an approach to automatically verify the queries' results. 
\parspaceskip{}


\normalpar{DBMS testing based on constraint solving}
Khalek et al. worked on automating testing DBMS using constraint solving~\cite{khalek2008,khalek2010querygen}.
Their core idea was to use a SAT-based solver to automatically generate database data, queries, and a test oracle.
In their first work, they described how to generate query-specific data to populate a database and enumerate the rows that would be fetched to construct a test oracle~\cite{khalek2008}.
They could reproduce previously-reported and injected bugs, but discovered only one new bug.
In a follow-up work, they also demonstrated how the SAT-based approach can be used to automatically generate queries~\cite{khalek2010querygen}.
As with our approach, they provide a test oracle, and additionally a targeted data generation approach.
While both approaches found bugs, our approach found many previously undiscovered bugs.
Furthermore, we believe that the simplicity of our approach could make it wider applicable.
\parspaceskip{}

\normalpar{Performance Testing}
Rather than trying to improve the correctness of DBMS, several approaches were proposed to measure and improve the DBMS optimizer's performance.
Poess. et. al proposed a template-based approach to generating queries suitable to benchmark DBMS, which they implemented in a tool QGEN~\cite{Poess2004}.
Similarly to random query generators, QGEN could also be used to test DBMS.
Gu. et al presented an approach to quantify an optimizer's accuracy for a given workload by defining a metric over different execution plans for this workload, which were generated by using DBMS-specific tuning options~\cite{Gu2012}.
They found significant accuracy differences for optimizers of multiple commercial database systems.



\section{Conclusion}
We have presented an effective approach for detecting bugs in DBMS, which we implemented in a tool \toolname{}, with which we found over \nrTruePositives{} bugs in three popular and widely-used DBMS.
The effectiveness of \toolname{} is surprising, considering the simplicity of our approach, and that we only implemented a small subset of features that current DBMS support.
There are a number of promising directions that could help uncovering additional bugs, which we regard as future work.
\toolname{} generates tables with a low number of rows to prevent timeouts of queries when multiple tables are joined with non-restrictive conditions.
By generating targeted queries with conditions based on table cardinalities~\cite{Bruno2006,Mishra2008}, we could test the DBMS for a large number of rows, better stressing the query planner~\cite{giakoumakis2008testing}.
Some language elements are difficult to test with our approach, for example, aggregate functions that compute results over multiple rows.
To efficiently test those, metamorphic testing~\cite{Chen1998} could be applied by defining metamorphic relations based on set operations.
Finally, we could also generate conditions and check that the \pivot{} is not contained in the result set, which might uncover additional bugs.

\section{Acknowledgments}
We want to thank all the DBMS developers for responding to our bug reports as well as analyzing and fixing the bugs we reported.
We especially want to thank the SQLite developers, D. Richard Hipp and Dan Kennedy, for taking all bugs we reported seriously and fixing them quickly.

\balance

\bibliographystyle{abbrv}
\bibliography{bib}  


\end{document}